\documentclass[10pt]{IEEEtran}
\onecolumn
\usepackage[ruled,vlined]{algorithm2e}
\usepackage{url,setspace}

\usepackage[table]{xcolor}
\usepackage[T1]{fontenc}
\usepackage{setspace,hyperref,enumitem}

\usepackage{array,graphicx}
\usepackage{amsfonts,cite}
\usepackage[cmex10]{amsmath}
\usepackage{mathtools}
\usepackage[amsmath]{empheq}
\usepackage{cases,balance,cite, lineno}
\usepackage{booktabs,mathrsfs,bbm,amssymb,soul}

\usepackage{wrapfig}
\usepackage{lscape}
\usepackage{rotating}

\usepackage{caption}
\usepackage{setspace}

\definecolor{myblue}{rgb}{0,0,1}

\long\def\aur#1{{\color{black}#1}} 
\long\def\ayu#1{{\color{black}#1}} 
\def\rev#1{{#1}}

\def\l{\left(}
\def\r{\right)}

\newcommand{\DEq}[1]{\stackrel{(\ref{#1})}{=}}

\def\DE{\stackrel{\mathrm{def}}{=}}

\def\Pb{(K^2-K)/2 + 1}
\newcommand{\bvec}[1]{\boldsymbol{#1}}
\newcommand{\ac}[1]{\mathcal{C}\left[ #1 \right]}

\begin{document}

\begin{center}
\par\noindent {\LARGE\bf Sampling Without Time: \\
Recovering Echoes of Light via Temporal Phase Retrieval
\par}
\end{center}

\begin{center}
{\vspace{4mm}\par\noindent
Ayush Bhandari$^\dag$, Aur\'elien Bourquard~$^\ddag$ and Ramesh Raskar$^\dag$
\par\vspace{2mm}\par}

{\vspace{2mm}\par\it
\noindent $^\dag$ Media Laboratory and $^\ddag$ Research Laboratory of Electronics \\ Massachusetts Institute of Technology Cambridge, MA 02139--4307 USA. \par}

 {\vspace{2mm}\par\noindent $\phantom{^{\dag,\ddag}}$\rm 
\textrm{ayush@MIT.edu \ $\bullet$ \ aurelien@MIT.edu \ $\bullet$ \ raskar@MIT.edu}
\par}

{\bf \vspace{8mm}\par\noindent
\color{black!50} To Appear in the Proceedings of IEEE ICASSP, 2017\footnote{The $42^{\rm{nd}}$ IEEE International Conference on Acoustics, Speech, and Signal Processing (ICASSP).}. \\
This work expands on the ideas discussed in \cite{Bhandari:2016b} and \cite{Bhandari:2016a}.
\par\vspace{4mm}\par}
\end{center}
{\vspace{6mm}\par\noindent\hspace*{5mm}\parbox{150mm}{
{\begin{center}\textbf{Abstract}\end{center}}
\emph{This paper considers the problem of sampling and reconstruction of a continuous-time sparse signal without assuming the knowledge of the sampling instants or the sampling rate. This topic has its roots in the problem of recovering multiple echoes of light from its low-pass filtered and auto-correlated, time-domain measurements. Our work is closely related to the topic of sparse phase retrieval and in this context, we discuss the advantage of phase-free measurements. While this problem is ill-posed, cues based on physical constraints allow for its appropriate regularization. We validate our theory with experiments based on customized, optical time-of-flight imaging sensors. What singles out our approach is that our sensing method allows for temporal phase retrieval as opposed to the usual case of spatial phase retrieval. Preliminary experiments and results demonstrate a compelling capability of our phase-retrieval based imaging device.} 
}\par\vspace{3mm}}
 \tableofcontents
 
 \newpage
 \begin{spacing}{1.5}
\section{Introduction}
\begin{figure}[!b]
    \centering
    \includegraphics[width =0.6\columnwidth]{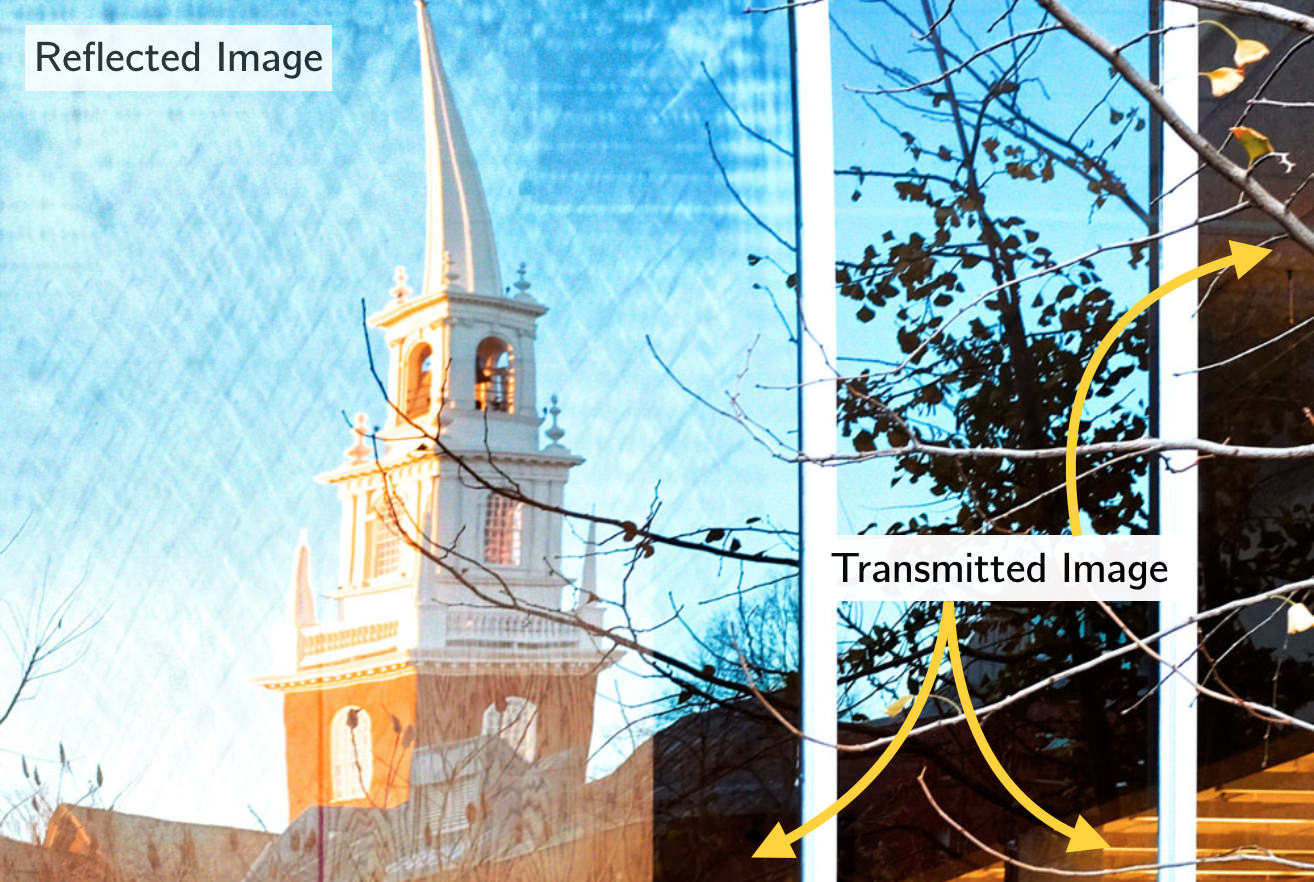}
    \caption{Reflection from semi-reflective surfaces. The Memorial Church can be seen imprinted on the glass facade of the Harvard University Science Center.}
    \label{fig:reflection}
 \end{figure}

During a recent visit to meet a collaborator, the author (who happens to be an avid photographer) saw a stark reflection of a local monument on the window panes of the \rev{Harvard} science center. This is shown in Fig.~\ref{fig:reflection}. This is a common place phenomenon at a macro scale as well as a micro scale (for example microscopy). At the heart of this problem is a fundamental limitation, that is, all of the conventional imaging sensors are agnostic to the time information. Alternatively stated, the image\rev{--}formation process is insensitive to the potentially distinct times that the photons can spend traveling between their sources and the detector.

\begin{figure}[!t]
    \centering
    \includegraphics[width =0.6\columnwidth]{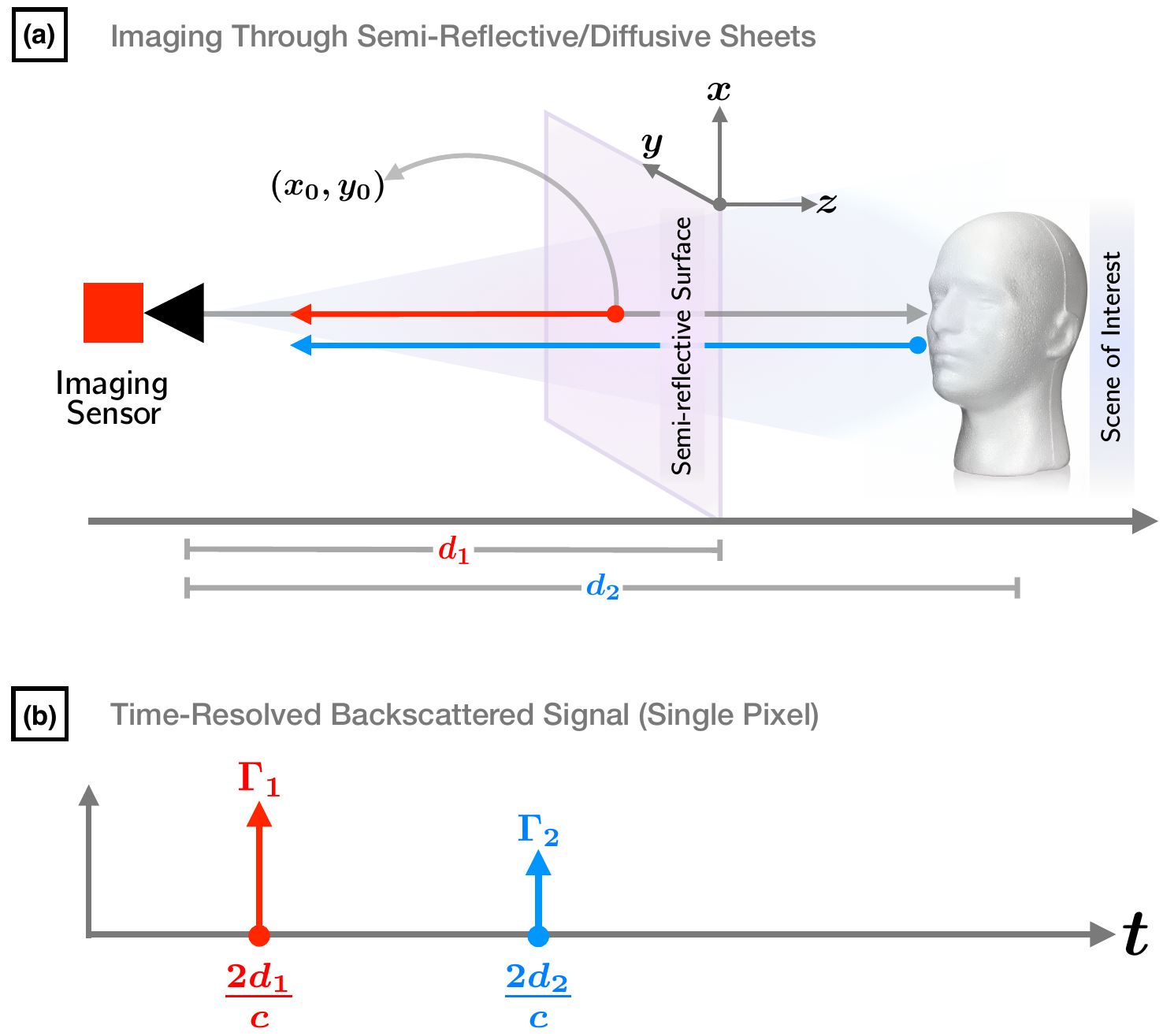}
    \caption{Time-aware imaging. (a) Exemplary setting for imaging through semi-reflective media. (b) Corresponding time-aware image at pixel $\bvec{r}_0 = \l x_0, y_0 \r ^{\top} $.}
    \label{fig:MPI}
 \end{figure}

To elaborate, consider the {\emph{Gedankenexperiment}} version of Fig.~\ref{fig:reflection} described in Fig.~\ref{fig:MPI} (a). Let us assume that a light source is co-located with the imaging sensor (such as a digital camera). The reflection from the semi-reflective sheet (such as a window pane) arrives at the sensor at time $t_1 = 2d_1/c$ while the mannequin is only observable on or after $t_2 = 2d_2/c$. Here, $c$ is the speed of light. While we typically interpret images as a two dimensional, spatial representation of the scene, let us for now, consider the photograph in Fig.~\ref{fig:reflection} along the time-dimension. For the pixel $\l x_0, y_0 \r$, this time-aware image is shown in Fig.~\ref{fig:MPI} (b). Mathematically, our time-aware image can be written as a $2$--sparse signal (and \rev{as} a $K$--sparse signal in general), 
\begin{equation}
m\left( {\bvec{r},t} \right) = \left( {{\lambda _1}{\Gamma _1}} \right)\left( \bvec{r} \right)\delta \left( {t - {t_1 {(\bvec{r})}}} \right) + \left( {{\lambda _2}{\Gamma _2}} \right)\left( \bvec{r} \right)\delta \left( {t - {t_2{(\bvec{r})}}} \right)
\label{TAI}
\end{equation}
where $\left\{ {{\Gamma _k}\left( \bvec{r} \right)} \right\}$ are the constituent image intensities, $\left\{ {{\lambda _k}\left( \bvec{r} \right)} \right\}$ are the reflection coefficients, $\bvec{r} = \l x , y\r^{\top}$ is the 2D spatial coordinate, and $\delta(\cdot)$ is the Dirac distribution.
%
%
%

A conventional imaging sensor produces images by marginalizing time information, resulting in the 2D photograph, 
\begin{equation}
\label{2DP}
\int_0^{\Delta  \gg {t_2}} {m\left( {\bvec{r},t} \right)dt}  = \left( {{\lambda _1}{\Gamma _1}} \right)\left( \bvec{r} \right) + \left( {{\lambda _2}{\Gamma _2}} \right)\left( \bvec{r}\right) \equiv I\l \bvec{r} \r.
\end{equation}
Recovering $\left\{ {{\Gamma _k}} \right\}, k = 1,2$ given $I$, more generally, $K$ echoes of light, 
\begin{equation}
\label{kmix}
m\left( {\bvec{r},t} \right) = \sum\limits_{k = 0}^{K - 1} {\left( {{\lambda _k}{\Gamma _k}} \right)\left( \bvec{r} \right)\delta \left( {t - {t_k\aur{(\bvec{r})}}} \right)}
\end{equation}
is an ill-posed problem. Each year, a number of papers \cite{Ohnishi:1996,Levin:2004,Bronstein:2005,Kong:2011,Li:2014,Chandramouli:2016} attempt to address this issue by using regularization and/or measurement-\aur{acquisition} diversity based on image statistics, polarization, shift, motion, color, or scene features. Unlike previous works, here, we ask the question: Can we directly estimate $\left\{ {{\Gamma _k}} \right\}_{k = 0}^{K - 1}$ in (\ref{TAI})? In practice, sampling (\ref{kmix}) would require exorbitant sampling rates and this is certainly not an option. Also, we are interested in $\left\{ {{\Gamma _k}} \right\}_{k = 0}^{K - 1}$ only, as opposed to $\left\{ {{\Gamma_k,t _k}} \right\}_{k = 0}^{K - 1}$ \cite{Bhandari:2016} where $t_k$ is a non-linear argument in (\ref{kmix}). As a result, our goal is to recover the intensities of a sparse signal. For this purpose, we explore the idea of \emph{sampling without time}---a method for sampling a sparse signal which does not assume the knowledge of sampling instants or the sampling rate. 
In this context, our contributions are twofold: 
\begin{enumerate}[leftmargin=40pt,label=$\arabic*)$]
\item For the general case of $K$-echoes of light, our work relies on estimating the constituent images $\left\{ {{\Gamma _k}} \right\}_{k = 0}^{K - 1}$ from the filtered, auto-correlated, time-resolved measurements. This is the distinguishing feature of our approach and is fundamentally different from methods proposed in literature which are solely based on spatio-angular information (cf.~\cite{Ohnishi:1996,Levin:2004,Bronstein:2005,Kong:2011,Li:2014,Chandramouli:2016} and references therein). 

\item As will be apparent shortly, our work is intimately linked with the problem of (sparse) phase retrieval (cf.~\cite{Eldar:2016,Qiao:2016,Rajaei:2016,Huang:2016,Lu:2011} and references therein). Our ``sampling without time'' architecture leads to an interesting measurement system which is based on time-of-flight imaging sensors \cite{Bhandari:2016b} and suggests that $K^2-K+1$ measurements suffice for \rev{estimation of} $K$ echoes of light. To the best \rev{of} our knowledge, \rev{neither such a measurement device nor bounds have been} studied in the context of image source separation \cite{Ohnishi:1996,Levin:2004,Bronstein:2005,Kong:2011,Li:2014}.
\end{enumerate}

For \aur{the sake of} simplicity, we will drop the dependence of $m$ and $\Gamma$ {on} spatial coordinates $\bvec{r}$. This is particularly well suited for our case because we do not rely on any \aur{cross-}spatial information or priors. Also, scaling factors $\lambda_k$ are assumed to be absorbed in $\Gamma_k$. 

\vspace{0.2cm}

\noindent{\bf Significance of Phase-free or Amplitude-only measurements:} It is worth emphasizing that resolving spikes from a superposition of low-pass filtered echoes, is a problem that frequently occurs in many other disciplines. This is a prototype model used in the study of multi-layered or multi-path models. Some examples include seismic imaging \cite{Levy:1981}, time-delay estimation \cite{Fuchs:1999}, channel estimation \cite{Barbotin:2012}, optical tomography \cite{Seelamantula:2014}, ultrasound imaging \cite{Boufounos:2011}, computational imaging \cite{Bhandari:2016b,Velten:2016,Shin:2016,Bhandari:2014a,Kadambi:2013} and source localization \cite{Fuchs:1994}. \aur{Almost} all of these variations rely on amplitude and phase information or amplitude and time-delay information. However, recording amplitude-only data can be advantageous due to several reasons. Consider a common-place example based on pulse-echo ranging. \rev{Let $p\l t\r = \sin \left( {{\omega}t} \right)$ be the emitted pulse. Then, the backscattered signal reads $r\left( t \right) = \Gamma_0\sin \left( {\omega t - {\theta _0}} \right)$.} In this setting, on-chip estimation of phase ($\theta_0$) or time delay ($t_0$)\cite{Bhandari:2014a},
\begin{enumerate}[leftmargin=40pt,label=$\bullet$]
\item can either be computationally expensive or challenging since $t_k$ is a non-linear parameter in (\ref{kmix}), and hence, in $r\left( t \right)$. 

\item requires more measurements. For instance, $2$ measurements suffice for amplitude-only estimation \rev{($\Gamma_0$)} while amplitude-phase pair requires $4$ measurements \cite{Bhandari:2016a}. This aspect of phase estimation is an important bottleneck as the the frame rate of an imaging device is inversely proportional to number time-domain measurements. 

\item is prone to errors. In many applications, multiple-frequency measurements, $\omega  = k{\omega _0}$, are acquired \cite{Bhandari:2014a,Bhandari:2015,Bhandari:2016a} assuming that phase and frequency, ${\theta _0} = \omega {t_0}$ are linearly proportional. However, this is not the case in practice. The usual practice is to oversample \cite{Hibino:1997,Hariharan:1987}.
\end{enumerate}
In all such cases, one may adopt our methodology of working with intensity-only measurements.

\section{Image--Formation Model for ToF Sensors}
Optical ToF sensors are active devices that capture 3D scene information. We adopt the generic image{--}formation model used in \cite{Bhandari:2016b} which is common to all ToF modalities such as lidar/radar/sonar, optical coherence tomography, terahertz, ultrasound, and seismic imaging. \aur{In its full generality, and dropping dependency on the spatial coordinate for convenience, one can first formalize this ToF acquisition model as:} 
\begin{equation}
p \to \boxed h \to r \to \boxed\phi  \to y\xrightarrow{{{\textsf{Sampling}}}}\bvec{y} \mapsto \underbrace {m =  \ac{\bvec{y}}}_{{\textsf{Intensity}}}
\label{IFM}
\end{equation}
where, 
\begin{enumerate}[leftmargin=40pt,label=$\rhd$]
\item $p\left( t \right) >0$ is a $T$-periodic probing function which is used to illuminate the scene. This may be a sinusoidal waveform, or even a periodized spline, \aur{G}aussian, or Gabor pulse\aur{, for instance}.
\item $h\l t , \tau \r $ is the scene response function (SRF). This may be a filter, a shift-invariant function $h_{\sf SI}\l t , \tau \r = h \l t - \tau\r$, or a partial differential equation \aur{modeling} some physical phenomenon \cite{Bhandari:2015}. 

\item $r\left( t \right) = \int {p\left( \tau  \right)h\left( {t,\tau } \right)d\tau}$ is the reflected signal \aur{resulting from the} interaction between the probing function and the SRF. 

\item $y\left( t \right) = \int {r\left( \tau  \right)\phi \left( {t,\tau } \right)d\tau }$ is the continuous-time \aur{signal resulting from the} interaction between the reflected function and the instrument response function (IRF), \rev{or $\phi$}, which characterizes the electro-optical transfer function of the sensor. 

\item $\bvec{y}$ is a set of discrete \aur{measurements of the form} ${\left. {y\left( t \right)} \right|_{t = n\Delta }}$. 

\item $\aur{\ac{\bvec{y}} = \bvec{y} * \overline {\bvec{y}}}$ is the cyclic auto-correlation \aur{of $\bvec{y}$,} where $*$ and $\overline {\l \cdot \r}$ denote the convolution and time reversal operators, respectively. 

\end{enumerate}
The interplay between $p$, $h$, and $\phi$ results in variations on the theme of ToF imaging \cite{Bhandari:2016b}. In this work, we will focus on \aur{an} optical ToF \aur{setting}. \aur{Accordingly:}

\begin{enumerate}[leftmargin=20pt,label=$\arabic*)$]
  \item The probing function corresponds to a time-localized pulse.

  \item The SRF, accounting for the echoes of light, is a $K$-sparse filter, 
\begin{equation}
\label{ksparse}
{h_{{\textsf{SI}}}}\left( t,\tau \right)\equiv h_K\l t - \tau \r = \sum\limits_{k = 0}^{K - 1} {{\Gamma _k}\delta \left( {t - \tau -  {t_k}} \right)}.
\end{equation}
  
  \item The IRF is fixed by design \aur{as} ${\phi _{{\sf{SI}}}}\left( t,\tau \right) = p\left( t + \tau \right)$. This implements the so-called homodyne, lock-in sensor \cite{Bhandari:2016b}. 
\end{enumerate}
Due to \aur{this specific} shift-invariant structure of the SRF and IRF, we have, $\bvec{y} = p*\overline p *{\overline h _K} \equiv \ac{p}*\overline{h} _K$.
Finally, the measurements read,  
\begin{equation}
\label{acm}
m = \ac{\bvec{y}}  = \l  \ac{\varphi}*\ac{{h_K}} \r, \quad \varphi = \ac{p}.
\end{equation}
 
\section{Sampling Echoes of Light} 

\subsection{Bandlimited Approximation of Probing Function}

\begin{figure}[!t]
    \centering
    \includegraphics[width =0.6\columnwidth]{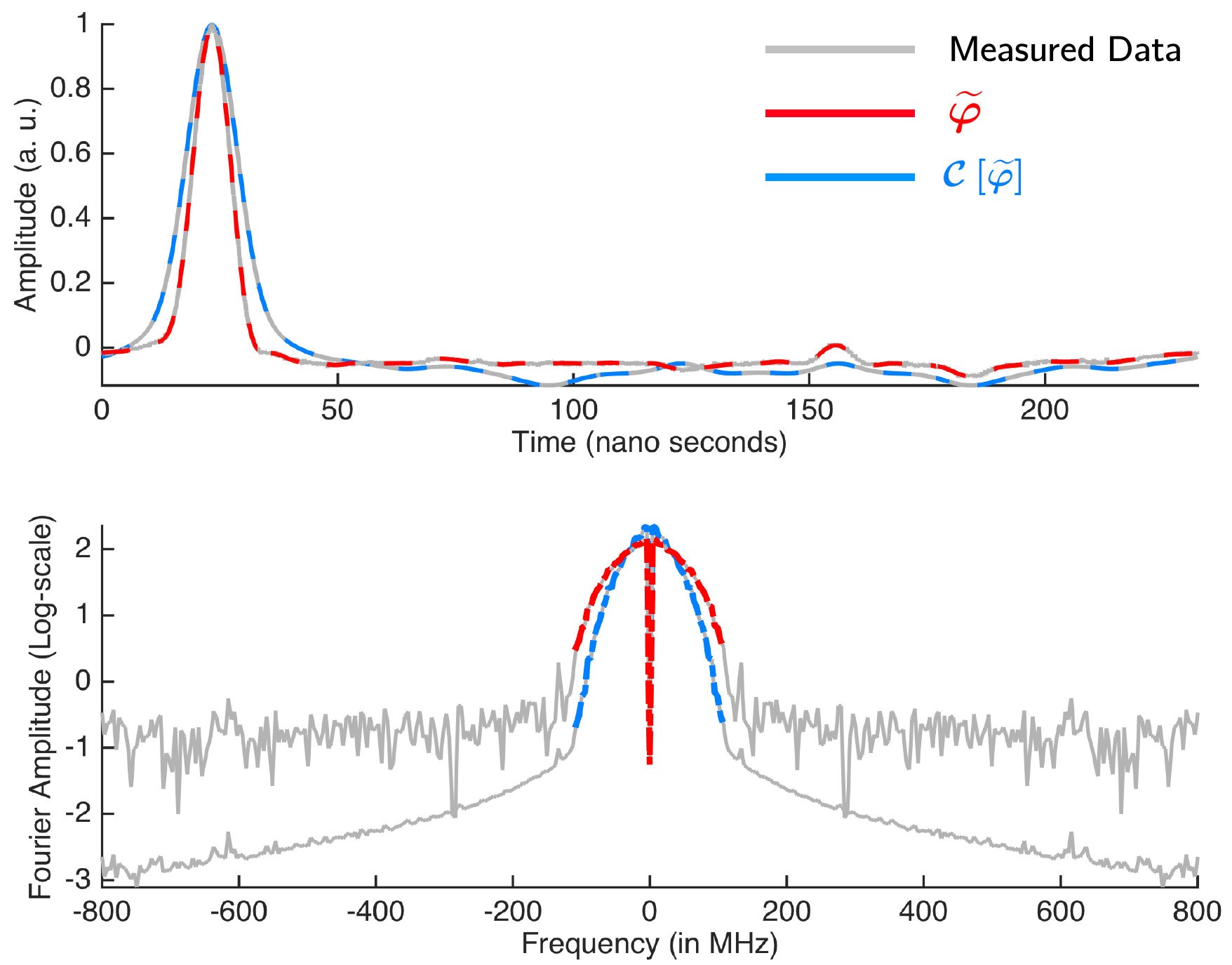}
    \caption{\aur{Bandlimited approximation $\widetilde\varphi \approx \ac{p}$ with $L = 25$ (red curves)} and \aur{its cyclic auto-correlation} $\ac{\widetilde\varphi}$ \aur{(blue curve)}. \aur{Shown are the spatial-domain signals (top) and the corresponding Fourier-domain magnitudes in log scale (bottom).}}
    \label{fig:fsa}
 \end{figure}

Due to physical constraints inherent to all electro-optical systems, it is reasonable to approximate the probing signal as a bandlimited function \cite{Bhandari:2015,Bhandari:2016}. We use $L$--term Fourier series with basis functions ${u_n}\left(\ayu{\omega_0 t} \right) \DE {e^{\jmath {\omega _0}nt}}$, $\omega_0 T = 2\pi$, to approximate $p$ with,
\begin{equation}
\label{FSA}
p\left( t \right) \approx \widetilde p\left( t \right) = \sum\limits_{\left| \ell \right| \leqslant L} {{{\widehat p}_\ell}{u_\ell}\left(  \ayu{\omega_0 t} \right)} , 
\end{equation}
where \aur{the} $\widehat{p}_\ell$ are the Fourier coefficients. Note that there is no need to compute $p\l t \r$ explicitly as we only require the knowledge of $\ac{p}$ (cf.~(\ref{acm})). In Fig.~\ref{fig:fsa}, we plot the calibrated $\ac{p}$ and its Fourier coefficients. In this case, $T = 232.40$ ns. We also plot its approximation, ${\widetilde\varphi} = \ac{\widetilde{p}}$ with $L = 25$ together with the measured $\ac{p}$ and $\varphi$.     

\subsection{Sampling Theory Context}
\aur{The} shift-invariant characterization of the ToF image\rev{--}formation model \aur{allows to re-interpret the} sampled version of (\ref{acm}) as \aur{the} filtering of a sparse signal $\ac{h_K}$ with a low-pass kernel $\psi = \ac{\varphi}$ (cf.~\cite{Pan:2017,Bhandari:2016}). 
\begin{align}
\label{samples}
m\left( t \right) & \DEq{acm} \ac{h_K}*\ac{\varphi} \equiv  \ac{h_K}*\psi. 
%
\end{align}
\aur{Note that the properties of the auto-correlation operation imply that} the sparsity of \aur{$\ac{h_K}$ is} $K^2 - K + 1$, \aur{unlike $h_K$ that is $K$-sparse} and is completely specified by $\Pb$ due to symmetry. \aur{Based on the approximation} (\ref{FSA}) \aur{and the properties of convolution and complex exponentials, and defining \aur{$\widehat\psi_\ell = |\widehat{p}_\ell|^4$}, we rewrite $m(t)$ as},  
\begin{equation}
\label{vmat}
\aur{m \l t \r} =  \sum\limits_{|\ell|\leq L} \widehat{\psi}_\ell u_\ell \l \ayu{\omega_0}t \r \underbrace{\int{\ac{h_K}\l \tau \r u_\ell^*{\l \ayu{\omega_0}\tau\r}} d\tau}_{\sf{Fourier\ Integral}},
\end{equation}
%
%
%
\aur{Finally, the properties of the Fourier transform imply that sampling} the above \aur{signal $m(t)$} at time instants $t = n\Delta$, results in \aur{discrete measurements of the form} $\bvec{m} = \mathbf{U} \mathbf{D}_{\hat{\psi}} \hat{\bvec{s}}$, \aur{which corresponds to the available acquired data in our acquisition setting. Combining all the above definitions, it follows that:}
\begin{enumerate}[leftmargin=40pt,label=---$\bullet$]
\item ${\bvec{m}} \in {\mathbb{R}^{N}}$ is a vector of filtered measurements (\aur{cf.} (\ref{samples})). 

\item ${\mathbf{U}} \in {\mathbb{C}^{N \times \left( {2{L} + 1} \right)}}$ is a DFT matrix with \aur{elements} ${\left[ u_n \l \ayu{\omega_0}\ell \Delta \r \right]_{n,\ell}}$.

\item ${{\mathbf{D}}_{\hat\psi}} \in {\mathbb{C}^{\l 2L+1\r \times \l 2L+1\r}}$ is a diagonal matrix with diagonal elements $\widehat\psi_\ell$. These are the Fourier\aur{-series} coefficients of $\ac{\varphi}$. 

\item $\widehat{\bvec{s}} \in \mathbb{R}^{2L+1}$ is \aur{a} phase-less vector \aur{containing the} Fourier coefficients of $\ac{h_K}$, which is obtained by sampling the Fourier transform $\widehat s\left( {{\omega }} \right)$ of $\ac{h_K}$ at instants $\omega = \ell\omega_0$. \aur{The signal $\widehat s\left( {{\omega }} \right)$ directly depends on the quantities $|\Gamma_k|$ of interest to be retrieved and is expressed as}
{
\begin{align}
\label{mag}
\widehat s\left( {{\omega }} \right) & = {{\left| {\sum\limits_{k = 0}^{K - 1} {{\Gamma _k}\ayu{u_{t_k}\l\omega\r}} } \right|^2}} \equiv |\widehat{h}_K \l \omega \r|^2  \hfill \\
& =  \underbrace{\sum\limits_{k = 0}^{K - 1} {|\Gamma _k|^2}}_{a_0} + 2\sum\limits_{k = 0}^{K - 1} {\sum\limits_{m = k + 1}^{K - 1} { \underbrace{\left| {{\Gamma _k}} \right|\left| {{\Gamma _m}} \right|}_{a_{k, m}} \cos \left( {{\omega}{t_{k,m}} + \angle {\Gamma _{k,m}}} \right)} }  \notag
\end{align}}
where ${t_{k,m}} = {t_k} - {t_m}$ and $\angle {\Gamma _{k,m}} = \angle {\Gamma _k} - \angle {\Gamma _m}$. 
\end{enumerate}
It is instructive to note that the relevant unknowns $\left\{ {{\Gamma _k}} \right\}_{k = 0}^{K - 1}$ can be estimated from $\widehat{\bvec{s}} \in \mathbb{R}^{2L+1}$ in (\ref{mag}) which in turn depend only on $L$ as opposed to sampling rate and sampling instants. This aptly justifies our philosophy of \emph{sampling without time}.

\section{Reconstruction via Phase Retrieval}

Given data model $\bvec{m} = \mathbf{U} \mathbf{D}_{\hat{\psi}} \hat{\bvec{s}}$ (\ref{vmat}), we aim to recover the image parameters $|\Gamma_k|$ at each sensor pixel. For this purpose, we first estimate samples of $\hat s(\omega)$ as $\widehat{\bvec{s}} = \mathbf{D}_{\psi}^{-1}\mathbf{U}^{+}\bvec{m}$, where $\mathbf{U}^{+}$ is the matrix pseudo-inverse of $\mathbf{U}$. This is akin to performing a weighted deconvolution, knowing that $\mathbf{U}$ is a DFT matrix. Next, we solve the problem of estimating $|\Gamma_k|$ in two steps (1) First we estimate \rev{$a_{k,m}$}, and then, (2) based \rev{on} the estimated values, we resolve ambiguities due to $|\cdot|$.\\

\noindent{\bf Parameter Identification via Spectral Estimation:} Based on the coefficients $\widehat{\bvec{s}}$, one can then retrieve the amplitude and frequency parameters that are associated with the oscillatory terms as well as the constant value in (\ref{mag}). The oscillatory-term and constant-term parameters correspond to $\{a_{k, m}, t_{k, m}\}$ and $a_0$, respectively. All parameter values are retrievable from $\widehat{\bvec{s}}$ through spectral estimation \cite{Stoica1997}; details are provided in \cite{Bhandari:2016a} for the interested reader.

Note that, given the form of (\ref{mag}) and our acquisition model, the sparsity level of the sequence $\widehat{\bvec{s}}$---corresponding to the total amount of oscillatory and constant terms---is $(K^2 - K)/2 + 1$. The magnitude values $a_{k, m}$ and $a_0$ can thus be retrieved if at least $L\geq (K^2 - K) + 1$ autocorrelation measurements are performed \cite{Potts:2010,Bhandari:2016a}, which is the case in the experiments described in Section \ref{Sect: Experiments}. 

\vspace{6pt}

\noindent{\bf Resolving Ambiguities:}
Based on the aforementioned retrieved parameters, one wishes to then deduce $\{|\Gamma_k|\}_{k=0}^{K-1}$. The estimated cross terms $\{a_{k, m}, t_{k, m}\}$ allow to retrieve the values of $|\Gamma_k|$ through simple pointwise operations. Here we will focus on the case, $K=2$. Due to space limitations, we refer the interested readers to our companion paper \cite{Bhandari:2016a} for details on the general case ($K>2$). The case when $K=2$, 
\begin{align*}
\widehat s\left( \omega  \right) \DEq{mag} & {\left| {{\Gamma _0}u_{{t_0}}^*\left( \omega  \right) + {\Gamma _1}u_{{t_1}}^*\left( \omega  \right)} \right|^2} + \underbrace {{\varepsilon _\omega }{{\left( {{\Gamma _k},{t_k}} \right)}_{k > 2}}}_{ \approx 0}\\
= &\underbrace {{{\left| {{\Gamma _0}} \right|}^2} + {{\left| {{\Gamma _1}} \right|}^2}}_{{a_0}} + \underbrace {2\left| {{\Gamma _0}} \right|\left| {{\Gamma _1}} \right|}_{{a_{01}}}\cos \left( {\omega {t_{0,1}} + \angle {\Gamma _{0,1}}} \right),
\end{align*}
may be a result of two distinct echoes {(Fig.~\ref{fig:MPI})} or due to approximation of higher order echoes by two dominant echoes (due to inverse-square law). In this case, we effectively estimate two terms: $a_0$ and $a_{01}$ (\ref{mag}). The set of retrieved magnitude values then amount to solving\cite{Bhandari:2016a}: 
\begin{equation}
\left| {\left| {{\Gamma _0}} \right| \pm \left| {{\Gamma _1}} \right|} \right| = \sqrt {{a_0} \pm {a_{01}}}, \qquad{a_0},{a_{01}} > 0.
\end{equation}
Thanks to the \emph{isoperimetric property} for rectangles: $a_0\geq a_{01}$, the r.h.s above is always positive unless there is an estimation error in which case, an exchange of variables leads to the unique estimates,
\[{\left\{ {|{{\widetilde \Gamma }_k}|} \right\}_{k = 1,2}} = \frac{{\sqrt {{{\widetilde a}_0} + {{\widetilde a}_{01}}}  \pm \sqrt {{{\widetilde a}_0} - {{\widetilde a}_{01}}} }}{2}.\]
\section{Experimental Validation}
\label{Sect: Experiments}
\subsection{Simulation}
Noting that measurements $\bvec{m} = \mathbf{U} \mathbf{D}_{\hat{\psi}} \hat{\bvec{s}}$ and $\hat{\bvec{s}}$ are linked by an invertible, linear system of equations, knowing $\bvec{m}$ amounts to knowing $\hat{\bvec{s}}$. We have presented detailed comparison of simulation results in \cite{Bhandari:2016a} for the case where one directly measures $\hat{\bvec{s}}$. In this paper, we will focus on practical setting where the starting point of our algorithm is (\ref{vmat}).

\begin{sidewaysfigure}[!t]
    \centering
    \includegraphics[width =1\textwidth]{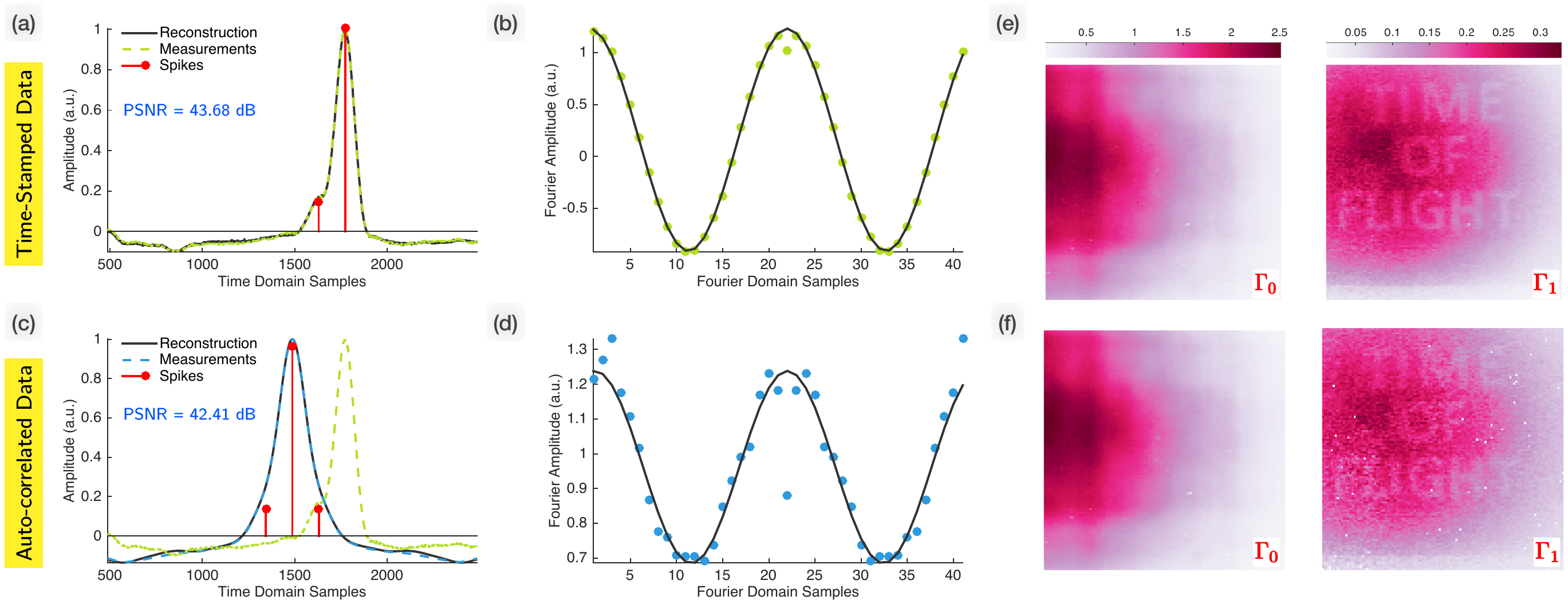}
    \caption{Experimental results on acquisition and reconstruction of echoes of light in a scene. (a) Single pixel, time-stamped samples $y = \ac{p}*\overline{h} _2$ serve as our ground truth because the phase information is known. We also plot the estimated, 2-sparse SRF (\ref{ksparse}) in red ink. (b) Fourier domain frequency samples of the SRF, $\widehat{h}_2 \l \ell \omega_0\r$ with $L = [-20,20]$ and $\omega_0 = 2\pi/\Delta, \Delta = 70$ ps. We plot the measured data in green and the fitted data in black. The frequencies are estimated using Cadzow's method. (c) Same pixel, time-stampless, low-pass filtered, auto-correlated samples (\ref{samples}) together with the estimated, auto-correlated SRF $\ac{h_2}$ in red ink. (d) Fourier domain samples $\widehat{s}\l \ell \omega_0 \r$ (\ref{mag}) and its fitted version (black ink). (e) Ground truth images for the experiment. (f) Estimated images using temporal phase retrieval.}
    \label{fig:mainresult}
\end{sidewaysfigure}

\subsection{Practical Validation}
Our experimental setup mimics the setting of Fig.~\ref{fig:MPI}. A placard that reads ``Time--of--Flight'' is covered by a semi-transparent sheet, hence $K = 2$. The sampling rate is $\Delta \approx 70 \times 10^{-12}$ seconds using a our custom designed ToF imaging sensor \cite{Bhandari:2016b}. Overall, the goal of this experiment is to recover the magnitudes ${\left\{ {|{\Gamma _k}|} \right\}_{k = 0}^{k=1}}$ given auto-correlated measurements $m\l n\Delta \r  \DEq{acm} \ac{y}\l n\Delta \r, \mathbf{m}\in\mathbb{R}^{2795}$. To be able to compare with a ``ground truth'', we acquire time-domain measurements $y\l n\Delta \r$ before autocorrelation, whose Fourier-domain phases are intact. In Fig.~\ref{fig:mainresult}(a), we plot the non-autocorrelated measurements $\mathbf{y}$ while phase-less measurements $\mathbf{m}$ are shown in Fig.~\ref{fig:mainresult}(c) from which we note that the samples are symmetric in time domain due to $m = \ac{y}$ (cf.~\ref{acm}). In the first case (cf.~Fig.~\ref{fig:mainresult}(a)) where $y = \ac{p} *\overline{h}_K$, we have \cite{Bhandari:2016b},
\[\widehat y\left( \omega  \right) = {\left| {\widehat p\left( \omega  \right)} \right|^2}h_K^*\left( \omega  \right),\quad {h_K}\left( \omega  \right) = \sum\limits_{k = 0}^{K - 1} {{\Gamma _k}{u_{{t_k}}}\left( \omega  \right)}.\]
Similar to $\mathbf{m}$ in (\ref{vmat}), in this case, the measurements read $\mathbf{y} = \mathbf{U}\mathbf{D}_{\widehat{\varphi}} \widehat{\mathbf{h}}$ and we can estimate the complex-valued vector $\widehat{\mathbf{h}}$. The phase information in $\widehat{\mathbf{h}}$ allows for the precise computation of $\{\Gamma_k,t_k\}_{k=0}^{K-1}$ \cite{Bhandari:2016b,Bhandari:2016}. These ``intermediate'' measurements serve as our ground truth. The spikes corresponding to the SRF (\ref{ksparse}) are also marked in Fig.~\ref{fig:mainresult}(a). %
Fourier-domain measurements $\widehat h_2^*$ linked with $y\l n\Delta \r$ are shown in Fig.~\ref{fig:mainresult}(b). Accordingly, ${\widehat h^*_2}\left( \omega  \right),\ \omega = \ell \omega_0, l = \{-L, \ldots, L\}$ where $\omega_0 = 2 \pi/\Delta$ and $L = 20$. We estimate the frequency parameters $\{\Gamma_k,t_k\}_{k=0}^{K-1}$ using Cadzow's method \cite{Cadzow:1988}. The resulting fit is plotted in Fig.~\ref{fig:mainresult}(b). 

In parallel to Figs.~\ref{fig:mainresult}(a),(b), normalized, autocorrelated data $\mathbf{m} = \ac{\mathbf{y}}$ are marked in Fig.~\ref{fig:mainresult}(c). The signal $\mathbf{y}$ is also shown as reference in green dashed line. We also plot the estimated $\ac{h_2}$ (autocorrelated SRF). In Fig.~\ref{fig:mainresult}(d), we plot measured and deconvolved vector $\widehat{\bvec{s}} = \mathbf{D}_{\psi}^{-1}\mathbf{U}^{+}\bvec{m}$ from which we estimate $a_0,a_{k,m}$ (\ref{mag}). The result of fitting using Cadzow's method with $41$ samples is shown in Fig.~\ref{fig:mainresult}(d).

The reconstructed images, $|\Gamma_{0}|$ and $|\Gamma_{1}|$, due to amplitude-phase measurements (our ground truth, Fig.~\ref{fig:mainresult}(a)) are shown in Fig.~\ref{fig:mainresult}(e). Alternatively, the reconstructed images with auto-correlated/intensity-only information are shown in \ref{fig:mainresult}(f). One can observe the great similarity between the images obtained in both cases, where only a few outliers appear in the phase-less setting. The PSNR values between the maps reconstructed with and without the phase information are of $30.25$ dB for $|\Gamma_1|$ and $48.88$ dB for $|\Gamma_0|$. These numerical results indicate that, overall, the phase loss that occurs in our autocorrelated measurements still allows for accurate reconstruction \rev{of the} field of view of the scene. 

Finally, in order to determine the consistency of our reconstruction approach, the PSNR between the original available measurements and their re-synthesized versions---as obtained when reintroducing our reconstructed profiles $|\Gamma_k|$ into our forward model---are also provided. As shown in Figs.~\ref{fig:mainresult}(a) and (c), the PSNR values in the oracle and phase-less settings correspond to $43.68$ dB and $42.41$ dB, respectively, which confirms that our reconstruction approach accurately takes the parameters and structure of the acquisition model into account. 
 \section{Conclusions}
In this paper, we have introduced a method to satisfactorily recover the intensities of superimposed echoes of light, using a customized ToF imaging sensor for acquisition and temporal phase retrieval for reconstruction. Up to our knowledge, this is the first method that performs time-stampless sampling of a sparse ToF signal, in the sense that we only measure the amplitudes sampled at uniform instants without caring about the particular sampling times or sampling rates. This innovation can potentially lead to alternative hardware designs and mathematical simplicity as phases need not be estimated in hardware.

\end{spacing}
\begin{spacing}{1.5}


\end{spacing}

\end{document}